# Magnetic Materials for Transcranial Magnetic Stimulation (TMS)

Max Koehler, Akshata Sangle, Stefan Goetz

## Abstract

Introduction: Various coils for transcranial magnetic stimulation (TMS) are widely available for clinical and research use. Surprisingly, these coils are almost all designed as air coils, which require large levels of energy to achieve a given magnetic flux density and in turn electric field strength, whereas in other sectors, such as power electronics or electrical machines, magnetic materials have been used for a long time to achieve higher efficiencies.

Objective: To test the impact on the electric and magnetic properties on an air coil of different (soft) magnetic materials. The test objects are various ferrite cores, laminated sheet materials of nonisotropic corn-oriented silicon-steel, non-oriented silicon-steel, as well as cobalt-iron, and soft magnetic compound powder cores with insulated particles.

Methods: We built a reconfigurable air coil in the form of a figure-of-eight coil with known electrical parameters, including sufficient inductance in case of saturation, which can be fitted with various magnetically conductive materials for a comparative analysis. The coil represents the typical conditions of TMS with a dominant but unavoidable magnetic gap for the brain. The increase in efficiency and the energy requirement for different materials and % MSO were determined in a series of tests. The increase in efficiency is based on pulse energy normalized tests and conclusions to the overall energy requirements are based on target field normalized tests. All test results are interpreted as absolute values and normalized to the coils (and cores) mass.

Results: Every material led to a reduction in coil current and voltage for the same target electric field strength. For the same field energy, every material yielded lower losses. Most common materials saturated already at very low currents. More material in thicker layers could shift the saturation point but at the cost of high weight.

Conclusion: Due to their low saturation flux density, ferrites appear unsuitable for the high amplitude requirements of TMS. Laminated sheet materials and powder cores reduce the pulse energy, but the laminated sheet material adds more weight for the same effect than powder cores. Thus, appropriate magnetic materials can reduce the required pulse energy. Saturation flux density is the most relevant parameter, whereas the permeability beyond a certain base level is practically irrelevant. Most importantly, the weight of a magnetic-core coil may always be increased compared to an air coil for the same target field.

# Introduction

Magnetic stimulation, such as transcranial magnetic stimulation (TMS), uses strong brief magnetic pulses to induce currents into tissue and activate neurons or muscles [1–6]. Said procedure requires substantial amounts of electrical energy. Furthermore, higher pulse energies are accompanied by higher losses, which contribute to the heating of the system and must be supplied by the charger. A major aspect of the high energy need is the magnetic flux generation, which is task of the stimulation coil. Therefore, optimization of stimulation coils has been a pivotal topic in the field. Initial intuitive design and experimental as well as computational improvements have recently led to mathematical optimization and projection to improve the efficiency of the system [7–9]. However, coil design, optimization, as well as known design concepts and rules are almost exclusively focusing on air coils.

In air coils, a large part of the magnetic flux and also the magnetic field is due to flux continuity unavoidably outside of the stimulation target and even the body. The share outside the body does not contribute to stimulation but contains a rather large share of the field energy. This high field and thus pulse energy does not itself constitute losses but requires high currents and high voltages, which lead to a whole list of issues: high losses and therefore increased coil heating due to the high current, shortening the timespan of uninterrupted treatment; danger to life due to the high voltage; challenges in device technology and high hardware cost of the pulse source, including semiconductors and capacitors, for controlling the high voltages and currents; and the large limitations of TMS devices, e.g., with respect to the pulse shape as well as high effort to change it [10,11].

Most technical devices that involve strong magnetic fluxes exploit magnetically conductive materials, such as electrical machines [12], inductors [13], or power transformers [14,15], to reduce flux leakage, magnetization energy, size, and loss. It is therefore remarkable that TMS does not, despite early academic prototypes and very few commercial products [16–25]. Commercially available coils are almost exclusively air coils and rather simple. Research has developed a good understanding and design rules to the level of full optimization for air coils. The understanding of coils with magnetic materials or even design parameters and material selection, however, is in comparison still rudimentary or entirely missing [18,19,22,26]. Furthermore, TMS deviates from other magnetic applications so that the material requirements are substantially different. However, there is little knowledge on important material properties and their quantitative influences.

To fill this gap, we compare 11 different materials with various thicknesses and material orientations, resulting in 44 different testing conditions. We compare the energy need for the same target electric field strength as well as the resulting losses (copper and magnetic), and the electrical properties of the coils. Furthermore, we referenced the properties to the mass as a major issue of iron-core coils, which affects practicality.

# Magnetic Conditions and Constraints in TMS

In magnetic stimulation, a strong current in a coil generates a magnetic field **H** in space. Dependent on the magnetic conductivity, also called magnetic permeability $\mu_r$ of the medium, the field in turn leads to a magnetic flux density **B** differentially, which forms a magnetic flux $\Phi = \iint \mathbf{B}(\mathbf{r}) \cdot \mathbf{da}$ [27]. The flux density is continuous, i.e., has no source no sink so that like a stream of water flux can neither vanish nor evolve but must form closed loops. The combination of the continuity of the magnetic flux and Ampere's law of generating magnetic fields through current, the generation of flux in TMS with a coil outside the head necessarily requires flux outside the head, which can tie up large shares of the pulse energy [28]. The stimulating effect is attributed to the electric field **E**, which is generated by the change of the magnetic flux density per Faraday's law, not the magnetic field.

The pulse energy is represented by the integral of the magnetic field energy density in space, given by $\iiint \mathbf{H} \cdot \mathrm{d}^3 r$ [27]. Whereas the flux and flux density are needed for a large stimulation effect, the magnetic field **H** represents the necessary driving force for a certain material to establish the flux and can be reduced through high permeability. Since noninvasively, magnetic materials can only be used outside the head and not be injected, all stimulation-effective magnetic flux loops will have a considerable path through materials with low permeability as tissue is magnetically widely unresponsive like air or free space. As the air and tissue path widely dominates, it is likely that covering large shares of the unavoidable path of the flux outside the body with materials with already moderate permeability, e.g., ten times that of free space, could substantially reduce the energy content, and higher permeabilities may lead to diminishing returns.

As the magnetic permeability is an electronic effect of certain otherwise unconstraint quantum factors, materials saturate if all angular momentums and spins of the limited pool of electrons with the right degrees of freedom have responded. For sufficiently large electric fields, TMS may require flux densities that substantially exceed typical levels in electric motors or transformers, likely deep in the saturation range of many materials. With a proper coil configuration, saturation can be delayed by increasing the cross section of the magnetic material to guide the same flux through a larger cross section for lower flux density without larger affecting the spatial electric field distribution in the brain.

Due to the high frequency content of TMS pulses in the kilohertz range in combination with the large flux densities, eddy currents in electrically conductive magnetic materials, such as iron, cobalt, nickel, and their unoxidized compounds may be critical. The eddy current losses grow with at least the frequency squared and flux density squared [29]. Thus, nonconductive materials (ferrites) or either very thin isolated sheets in laminates with the right orientation or rather small isolated powder particles in a nonconductive polymer matrix may be required to keep the loss under control.

## Coil and Materials

We use a purpose-designed figure-of-eight base coil, which allows the addition of magnetic material on the back for flux guidance and field energy reduction. This design is topologically equal to C-core coils but keeps the proven shape of a figure-of-eight coil including the handling or compatibility with periphery and navigation. It can guide up to half of the magnetic flux through magnetic material on the back of the coil and has a clearer saturation behavior than C-core shapes as it merely degrades to an air coil. Furthermore, in contrast to a C-core coil, which fixes the magnetic material cross section as an early design parameter, the figure-of-eight setup with back material allows a flexible increase of the magnetic material layer at the cost of additional weight for testing all materials in the same design and avoiding individual designs with different focality.

For reproducibility and equal conditions for material specimens, we manufactured the base figure-of-eight coil with replaceable magnetic core. We wound the coil from 10 AWG wire (Alpha Wire 391045, Elizabeth (NJ), USA), placed it at the bottom of a polymer case, and fixed the winding block with a compound epoxy–titanium-dioxide (64.4 % epoxy resin, 25.9 % hardener, 9.5 % titanium-dioxide) (Fig. 1). The coil has 2 × 9 turns with the smallest turn starting in the center. The outset turn was 100 mm × 100 mm. The bottom layer of the coil casing has a thickness of 1.8 mm. The winding block has a thickness of 10 mm so that the magnetic core on the back of the winding block would reside about 12 mm above the coil-scalp surface. The wire outer overall outer diameter was 5.3 mm. The coil weighs 788 g, not including the connecting wires. The casing forms a rectangular volume with 100 mm × 200 mm footprint for specimens of various thicknesses up to 40 mm. The inductance of the coil without magnetic material was designed to the lower end of TMS coils (measured at 9.25 µH, Hameg Instruments

HM8118, Frankfurt, Germany) to operate also as air coil and protect the pulse source even in deep saturation and subsequent collapse of the inductance to this air level. A commercial MagVenture Mag-Lite pulse source (Tonica, Farum, Denmark) served as the pulse source for all tests.

Materials were selected with specific conditions in mind: most obviously the magnetic permeability $\mu_r$, which represents the magnetic conductivity of a material, typically determines how much energy a certain magnetic flux requires so that we chose materials from a range of low to high values; due to the high flux densities of up to 2 Tesla, we selected materials from a wide range of saturation levels; and in response to the high frequency, which exceeds typical power and motor applications but falls short of transformers in recent power-electronic isolation, we picked materials with various levels of electrical conductivity and high-frequency losses.

The selected materials belong to three main material groups: solid ferrites, soft-magnetic compound powder cores with insulating polymer binder and laminated magnetic sheets of various iron alloys (silicon-steel, cobalt-iron). Table 2 lists the materials with their physical properties, such as saturation flux density, relative permeability, mass density and material constellation. Ferrites are, because of their low electrical conductivity and therefore their low eddy current losses, often used in high frequency applications, but typically saturate notably below the Tesla range. Powder cores and laminated magnetic steel sheets on the other hand tolerate higher flux densities but can generate large losses at high frequencies. For laminated sheets, the direction of the sheets influences the magnitude of eddy currents [18] . We investigated the impact of this effect on various materials by aligning the laminated sheet stacks in different ways.

The ferrites and the soft-magnetic compound materials from Magnetics fit into the coil without modification. Other materials were supplied in various shapes, such as Höganäs Somaloy or Hitachi AMCC, which were machined and assembled from smaller elements to fit the 100 mm × 200 mm dimensions of the test coil. The laminated magnetic material samples were manufactured in-house from larger sheets cut to size (guillotine shears) and laminated together with bonding varnish (Backlack, EB549, Rembrandtin, used for Vacodur49, AMCC and Powdercore; Backlack, 1175W/K, Axalata, used for NO20-13) to form stacks of various size to represent three different lamination orientations (long, short and mixed, Table 1). Two different bonding varnishes were used as the NO20-13 sheets came pre-applied with bonding varnish from the manufacturer.

Since the datasheets of the Höganäs materials and the regular datasheet of Neosid PFS3 do not provide sufficient information about the material composition, we performed several analysis methods to fill this gap of knowledge and provide the data to scientific community.

## Characterization Methods

We sampled the electric field in the region of interest of every configuration from 0 % to 100 % machine output (MSO) with an automated field probe further described in the literature [30]. The electric field was sampled at a distance of 20 mm below the coil surface, which represents a typical coil-to-cortex distance. Concurrently, we measured the coil current (PEM CWT 60, Nottingham, UK) as well as the stimulator voltage (MICSIG DP7001, Shenzhen, CN). We quantified the field strength through the integral of the biphasic pulse between the first and the second zero crossings of the cosine electric field pulse instead of only the field peak, which occurs around the zero-crossing of the current and would therefore over-estimate the effect in case of saturation [31–33]. Furthermore, we measured the electrical properties of the different material combinations in magnitude and phase over the frequency spectrum from 20 Hz to 200 kHz with an amplitude of 1 V, i.e., below saturation (Hameg Instruments HM8118).

We generated Bode plots and additional extracted the arcus tangent of the phase to better characterize the behavior near the –90° mark (mere inductance, no resistance) to identify material-dependent losses. We furthermore extracted the target electric field (20 mm below the coil, center) over losses per pulse at 100% MSO, electric field per loss in relation to the coil weight, energy and loss reduction for normalized pulses at the electric field strength of the air coil at 100% MSO, electric field per loss and electric field over weight at target field strength and saturation effects of the materials. We further detected saturation in the pulse length, the effective inductance at each coil current $L(t) = \frac{u(t)}{\frac{di(t)}{dt}}$ occurring initially at the current maxima (close to the voltage zero-crossings, which are shifted dependent on the loss), and the induced electric field over the respective current readings.

We analyzed the materials Neosid PFS3, Neosid F02, Höganäs Somaloy 130i 1P 600, and Höganäs Somaloy Prototype with energy-dispersive X-ray (EDX) spectroscopy for material composition (Bruker XFlash Detector 630M, Billerica, MA, USA) and scanning electron microscopy (SEM, JEOL JSM-6610LV, Freising, Germany) at an acceleration voltage of 30 kV for microstructure imaging and grain size analysis. The procedure first prepared suitable samples of each material which we polished to achieve a smooth representative surface. We acquired SEM images to study and examine the microstructure for features like particle size, shape, porosity and interparticle structure. We recorded EDX spectra by targeting the electron beam at specific points on the samples to stimulate the emission of characteristic X-rays for material identification.

## Results

Some of the figures only include the key conditions. The supplement presents the complete set of data. The saturation flux density of powder cores and sheet materials exceeded the saturation flux density of the ferrites, whereas the ferrites and the sheet materials had higher magnetic permeabilities than the powder cores (Table 2 and Figure 2). As a direct result of low magnetic saturation flux density, the materials from the category ferrites have the lowest saturation flux density per mass density. Their lower mass density cannot compensate the lower absolute saturation flux density. The materials from Höganäs, Vacuumschmelze (VAC), and ThyssenKrupp therefore appear favorable in saturation density per mass density. Laminated sheet materials introduce further difficulties, examined below.

Generally, every material combination showed expected resistive–inductive behavior with minor differences (Figure 3). Neosid PFS3 (material sample no. 9) is almost in line with the air coil in amplitude as well as phase, although the material is thin (1.4 mm) and has the lowest permeability of all tested materials. The samples of the materials from Magnetics demonstrate a higher impedance with near zero phase and thus effective resistance already below 120 Hz, which represent loss. The impedances of all other materials and conditions demonstrate stronger inductive behavior with early phase transitions. However, only Neosid PFS3 (material sample no. 9) demonstrates the largest desired inductive properties with low loss at higher frequencies as the arcus tangent plot discloses. The remaining materials can be ordered as follows with respect to high-frequency losses (from lowest to highest): Ferrites, powder cores from Magnetics, Hitachi AMCC Series, Wälzholz NO-20, ThyssenKrupp Powercore, Höganäs Somaloy, and Vacuumschmelze Vacodur49.

Advertised as a *plastoferrite* without further definition, the SEM and EDX analysis revealed that Neosid PFS3 fulfills the definition of a powder core (Fig. 4). It consists of FeSi$_3$ particles bound together with a thermoplastic polymer (poly-oxymethylene). The material safety data sheet supports the interpretation. The particle size reaches up to 10 µm (Figure 4).

Both Höganäs materials consist of iron particles varying in size from 20 µm to 100 µm for Somaloy 130i 1P 600 as well as 100 µm and closely above for Somaloy Prototype Material (see SEM images in Fig. 3). Personal communication with an expert suggests that every Höganäs Somaloy material consists of isolated iron particles pressed together below sintering pressure [34–39]. In contrast to other powder materials, which use typically connect and isolates the grains with a polymer matrix, Höganäs preferably isolates the grain through chemical coating such as oxidation or phosphating. The Höganäs materials primarily seem to differ in their purity of the iron, grain size, insulation quality, and compression force, whereas compression force and insulation quality define the core losses. The Somaloy Prototype Material, which is closest to Somaloy 700 1P 600, furthermore includes additives for improved manufacturing [39].

Figure 5 illustrates the electric field in the target (20 mm below the coil, as above) per pulse power loss at 100% MSO. The red lines indicate constant ratios of field strength per loss. Starting with the leftmost line at the ratio of air, the lines to the right indicate progressively improving ratios. The first segment between the ratio of air and a ratio of $p = 0.13 \frac{\text{V} \cdot \text{s}}{\text{J} \cdot \text{m}}$ contains every laminated sheet material with the orientation *short* and *mixed* (19, 21, 23, 30, 33, 35, 37, 39, 41, 43), which performed worst across the board. Its ratio of target electric field per loss was slightly above the air coils ratio. In addition, all ferrites as well as the thick Kemet samples (2, 3, 4, 5, 6, 8, 9), thin Magnetics samples (17), and the thinnest laminated sheet material samples with the orientation *long* (20, 30, 34, 40) populate this segment. The thick ferrite (7), medium laminated sheet material in the *long* orientation (22, 24, 27, 28, 29, 32, 36, 38, 42, 44) and mixed (32), medium powder cores from Magnetics (15, 16, 18) and from Höganäs (10, 11, 12) offer intermediate target electric field to loss ratios (between $p = 0.13 \frac{\text{V} \cdot \text{s}}{\text{J} \cdot \text{m}}$ and $p = 0.15 \frac{\text{V} \cdot \text{s}}{\text{J} \cdot \text{m}}$. Thick powder cores from Höganäs (13, 14) and thick laminated sheet materials (25, 26) form the high-performance sector ($p = 0.15 \frac{\text{V} \cdot \text{s}}{\text{J} \cdot \text{m}}$ and above). The pulse power loss is relevant for the sizing of the capacitor charger in a TMS device, the power it draws from the outlet, and also the coil heating. The latter further scales with the effective thermal mass. All magnetic materials reduce the coil current so that their overall losses of all material configurations outcompete the air coil, some more or less. The additional loss mechanisms in the cores, such as hysteresis, do not come close to saved Ohmic loss due to the lower current in the winding.

Figure 6 relates the energy loss to the electric field in the target and plots it over the mass of coil. The powder cores from Höganäs yield the best overall results. For most materials, particularly the Somaloy Prototype Material, the gain in in electric field per loss does not linearly increase with the layer thickness and therefore the incorporated material mass but flattens sub-linearly. Closest in performance to the Somaloy Prototype Material are the Magnetics KoolMu cores (17,18), followed by the Waelzholz NO20-13 as a standard silicon-steel material. Vacuumschmelze Vacodur49 (cobalt-iron) follows the trajectory of Waelzholz NO20-13 closely but starts to noticeably drop in performance after 3.5 kg. ThyssenKrupp Powercore (non-grain-oriented silicon-steel with larger sheet thickness than Waelzholz NO20-13, which let expect increased eddy-current loss) consistently remains below the performance of the other sheet materials. Ferrites are situated at the bottom below the sheet materials. The Magnetics XFlux cores follow a different trajectory as they have a higher field per loss ratio than NO20-13 at 2.5 kg and a lower field per loss ratio than NO20-13 at 4.3 kg.

Figure 7 graphs the ratio of the electric field in the target per energy loss per coil mass and parallels it with the electric field per mass. Thus, both penalize the results for practically in TMS. None of the magnetic material configurations can outperform the sole air coil in either target field per loss or even target field per se when the weight is considered.

Figure 8 further compares the required pulse energy to reach the air coil's target electric field strength at 100 % MSO and the corresponding pulse loss. As the trend in Figure 3 already foreshadowed, the steel sheet material in the lamination direction *short* performed worse than the sheet material laminated in the direction *long* at the same thickness and correspondingly the same weight as it required more initial pulse energy and had higher losses. Sheet materials in the lamination direction *short* were able to reduce the required pulse energy to 84 % – 72 % and the energy losses to 87 % – 74 %, whereas the materials with the lamination direction *long* were able to reduce the required pulse energy to 70 % – 50 % and the energy losses to 69 % – 47 %. This direct comparison reveals that even the worst performing material in the direction *long* is able to beat the best performing material in the direction *short*. The powder cores achieved a reduction in field energy to 63 % – 47 % and a reduction in energy losses to 59% – 43 %. Ferrites decreased the required pulse energy to 93 % – 59 % and the energy losses to 92 % – 55 %.

Figure 9 further normalizes the pulse energy and pulse loss to the coil weight. Material samples no. 8 and 9, which performed poorly when comparing the absolute values from Fig. 7, show significant improvements for energy requirements, yet still unable to exceed the air coil. In conclusion, no material offers sufficient pulse energy or power loss savings to allow a lower coil weight.[1] All materials fall short by factors 1.16 to 4.52 in pulse energy per coil mass and 1.16 to 4.65 in pulse loss per coil mass in direct comparison with the air coil.

As expected, we obtained substantial magnetic saturation effects, especially from the materials with a lower saturation flux density. The effect was most pronounced for ferrites, especially at lower material thicknesses. The saturation manifested as a rather sudden drop in the electric field when the material saturates (Figure 10). Across all pulses for a sample, saturation set in at almost exactly the same current and thus flux level (Figure 11). For the same material, the saturation current level scales widely linearly with the material thickness as the flux can spread across a larger cross section for thicker layers. Thus, as expected, a low saturation flux density can be compensated by a thicker material backing of the coil and larger weight. The Kemet FPL material saturated at 500 A for a thickness of 8 mm, 1000 A for 16 mm, and 2000 A for 32 mm.

With saturation, the effective coil impedance dropped to the level of the air coil (Figure 12). At a thickness of 8 mm, the saturation in Kemet FPL sets in at very low currents from $14\ \mu H$ to $10\ \mu H$ at 3000 A. The curves shift with material thickness to a saturation current of 1300 A for a thickness of 32 mm, starting at $14\ \mu H$ and dropping off to $10{,}5\ \mu H$ at 6000 A. The decreased inductance determines the resonance frequency of the LC-oscillator circuit formed by the coil and the stimulator according to $T = 2\pi \cdot \sqrt{LC}$ and shortens the pulses with increasing pulse strength (Figure 13).

Figure 14 visualizes the nonlinear relationship between coil current and electric field strength. The dashed black line extrapolates the initial gradient and represents a hypothetical behavior of a coil without saturation. The colored lines indicate the actual measurements of the Kemet FPL Series ferrites. Even at a thickness of 32 mm, the saturation current level is far below maximum machine output levels. Every material tested demonstrated decreasing permeability of full saturation for the tested thicknesses, but ferrites failed the earliest.

---

[1] The results do not support component-level savings, i.e., on the level of the coil, the results cannot inform potential system-level savings as those would involve very specific electronics design concepts and decisions.

## Discussion

The most important material property for magnetic cores for TMS coils was saturation flux density per mass density and least relevant the magnetic permeability as long as it was in the double digits. Materials with higher saturation flux density may demonstrate larger high-frequency losses (Figure 3), particularly ferrites versus iron materials. However, for TMS pulses (typically 2.5 to 7 kHz carrier frequency), and strength, a high saturation level still turned out more important.

Thus, even though all ferrites performed well at high frequency with minimum loss, their relatively low saturation flux density was least suitable for TMS. They reduced the required pulse energy by 7 % to 41 % and the energy losses by 8 % to 45 % for the same target electric field. The powder core materials and most of the laminated steel sheet materials performed better (overall and per weight) than the ferrites. The ferrites further demonstrated a strong magnetic field and thus voltage increase in the event of saturation, which strongly distorted also the electric-field pulse shape. The quantity of material needed to entirely prevent saturation and the associated pulse shape distortions at 100 % MSO can be linearly extrapolated to a core mass of 11.2 kg (compared to typical coil masses of up to 2 kg for most commercial figure-of-eight coils), which is beyond manually manageable weights and also exceeds the typical 5 kg limit of most TMS robots at the moment [40–44]. Since a combined mass of 12 kg is neither manageable in a handheld application nor through most robotic actuators, ferrites prove to be unsuitable for the application of TMS.

The lamination direction of the sheet materials had substantial influence as theoretically suggested previously [18]. Lamination such that the flux lines impact the sheets perpendicularly (direction *short*) caused higher losses and a notably lower performance, i.e., more losses per field (due to the many material interfaces for the flux from sheet to sheet through the magnetically poor varnish) than their counterparts with the sheets along the field lines (lamination direction *long*). We therefore ruled out samples no. 19, 21, 23, 30, 33, 35, 37, 39, 41, and 43.

The most suitable material for TMS were powder cores. Höganäs Somaloy Prototyping or 130i 1P 600 outperformed the other materials in one or more aspects: The ratio between field strength and losses was highest, especially in comparison with other samples of comparable weight (Figures 4 and 5). They were easiest to manipulate (machining or even glueing). Ferrites in contrast were brittle and tended to break when machined. Laminated steel alloy sheets used established processes from electric motors, but tooling and lamination can be expensive for the typical quantities of TMS. The studied powder core materials were widely magnetically isotropic so that their orientation was irrelevant in contrast to the laminated sheet cores.

The powder cores offered high magnetic saturation flux densities, which were the limiting factor for ferrites. The saturation flux density of the Höganäs materials were 1.4 times higher than of standard silicon-steel, 1.15 times higher than the best lamination (cobalt-iron), and 4.4 times higher than of ferrites. Since the mass density is on par with steel and only about twice that of ferrites, the powder material had the highest saturation flux density per material density. Thus, the material required the lowest material weight to guide a given magnetic flux.

The powder materials had the lowest magnetic permeability (up to ten times lower than ferrites and up to 150 times lower than the laminations). In contrast to initial intuition, the low permeability was not relevant. As about half of the magnetic flux path is in air or tissue, i.e., with a relative permeability of 1, values above ten already unlocked more than 90 % of the potential of a magnetic core and were sufficient. TMS is rather different to other applications of technical magnetics, which either have no or only very small air gaps [45–48]. In TMS with a noninvasive coil, a long path through low-$\mu$ tissue is unavoidable with the current physical principles and paradigms.

The material thickness is the main determinant of the coil weight and may depend on the application. In cases that need the lightest possible coil not considering cooling, cable, or pulse source size, no material could compete with air coils. The addition of a 10 mm layer of Somaloy Prototype Material increased the mass by 1.3 kg and reduced the required pulse energy to 63.5 % and the losses to 58.9 %. If a fixture or a robotic actuator allow for higher masses, a 5.7 kg core can decrease the pulse energy to 47.9 % and the energy losses to 43.2 %. These numbers may inform a system-level weight optimization to overcompensate the coil weight: the reduced pulse energy reduces the capacitor and power-semiconductor size as well as the cable cross section.[2] The lower losses in turn can shrink the power supply. Importantly, the energy density in capacitors at TMS conditions are similar (~200 J/kg) than in air coils higher than in the magnetic-core coils due to the air or brain gap so that system-level optima through magnetic materials require tight weight budgeting and appear not trivial to reach.

The key determinants are the mass of the material and the large brain gap. Substantial gains appear achievable with a material with low weight per saturation flux density at the cost of a low relative permeability above 10 in combination with methods to reduce the air and brain gap in the flux path.

If lowest coil weight is desired with typical magnetic materials, contraire to prior findings [49], an air coil appears to offer the best choice. As the weight directly penalizes the stimulation field strength and the performance of the coil, a performance-doubling core is only allowed to weigh 800 grams. The closest contender to the air coil was a high-saturation ferrite / soft magnetic compound, specifically Neosid PFS3 (sample no. 9). A back layer of 1.4 mm adds only 144 g to the mass. The high saturation flux density of ferrites at 1 Tesla and the lower mass density outperformed all other materials relative to material weight. However, it still reached only 85 % of the benchmark set by the air coil in energy-normalized testing (Figure 7) and 86% of the benchmark in field-normalized testing (Figure 9). Despite good results with respect to weight, sample no. 9 performed poorly in other metrics. At the matched pulse strength of 100 % MSO air coil, sample no. 9 leads to a reduction to only 98.2 % of the required pulse energy and reduced energy losses to only 97.9 %, which places it last in the ranking (Figure 8). If compared to other materials at 100 % MSO, it again ranks last with an electric field to loss ratio of $p = 0.1048 \frac{\text{V} \cdot \text{s}}{\text{J} \cdot \text{m}}$ (Figure 5), Thus, it is almost indistinguishable from air.

# Acknowledgements

The authors are inventors named on patents and patent applications for brain stimulation technology independent from this paper. No external funding was received in conjunction with the research findings presented in this article.

---

[2] If the reduction of pulse energy is used to reduce the pulse current for equal voltage.

# Figures and Tables

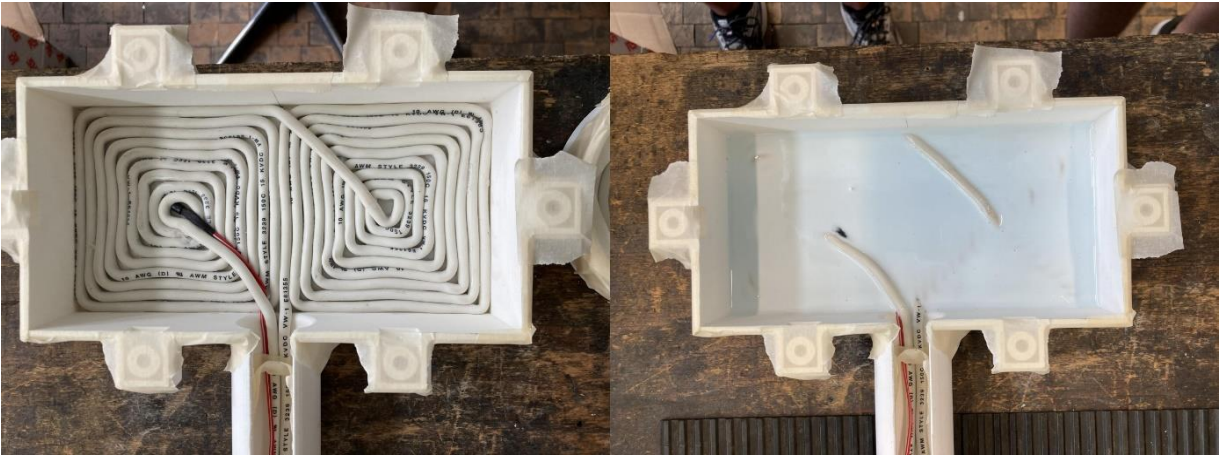

**Figure 1.** Left: Wiring of the test coil in progress in a custom-made casing with inner dimensions of 100 x 200 mm and bottom thickness of 1.8 mm, which can hold the material samples reproducibly. Right: Molded test coil with a mixture of epoxy resin and titanium oxide.

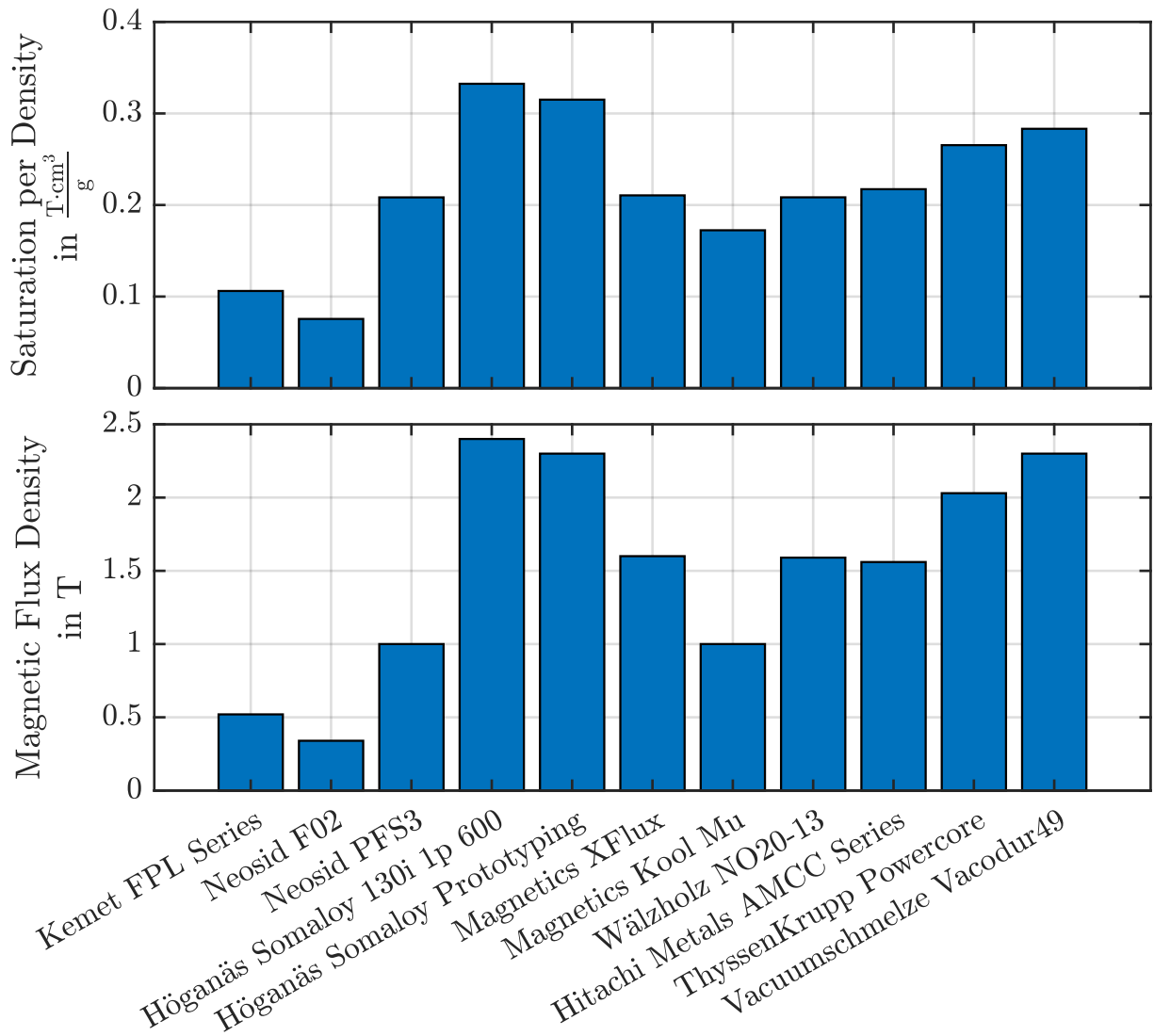

**Figure 2.** The top plot shows the saturation flux density divided by the Material density of each material. The bars indicate the maximum flux that can be guided per material weight (higher is better). The lower plot indicates the magnetic saturation flux density of each material.

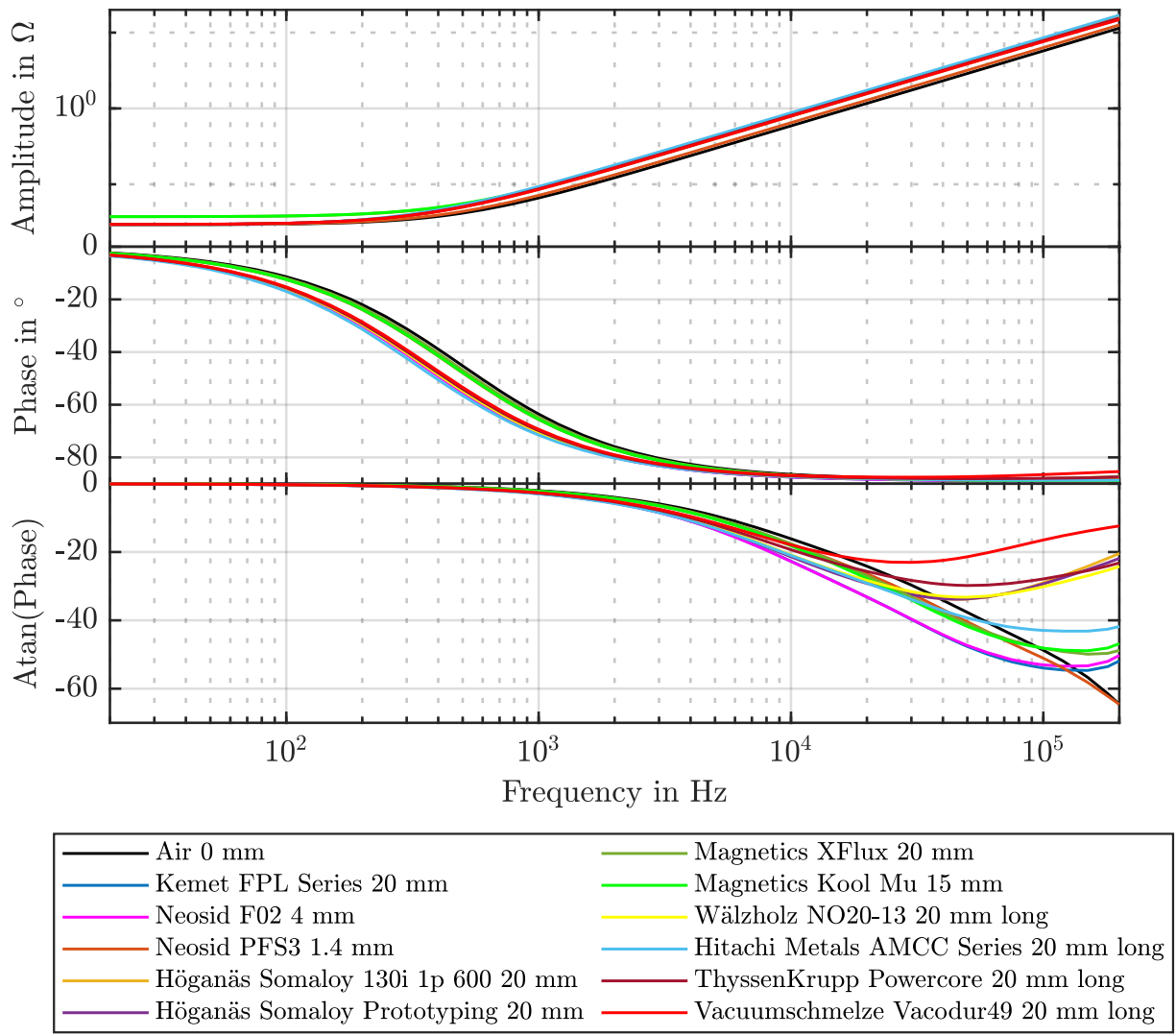

**Figure 3.** Bode Plot of every practical combination of materials. The top panel shows the amplitude of the system in Ω. The middle panel plots the phase in degree. The bottom panel represents the phase of the systems as its arctangent to point out the ratio of inductance relative to loss-representing resistance.

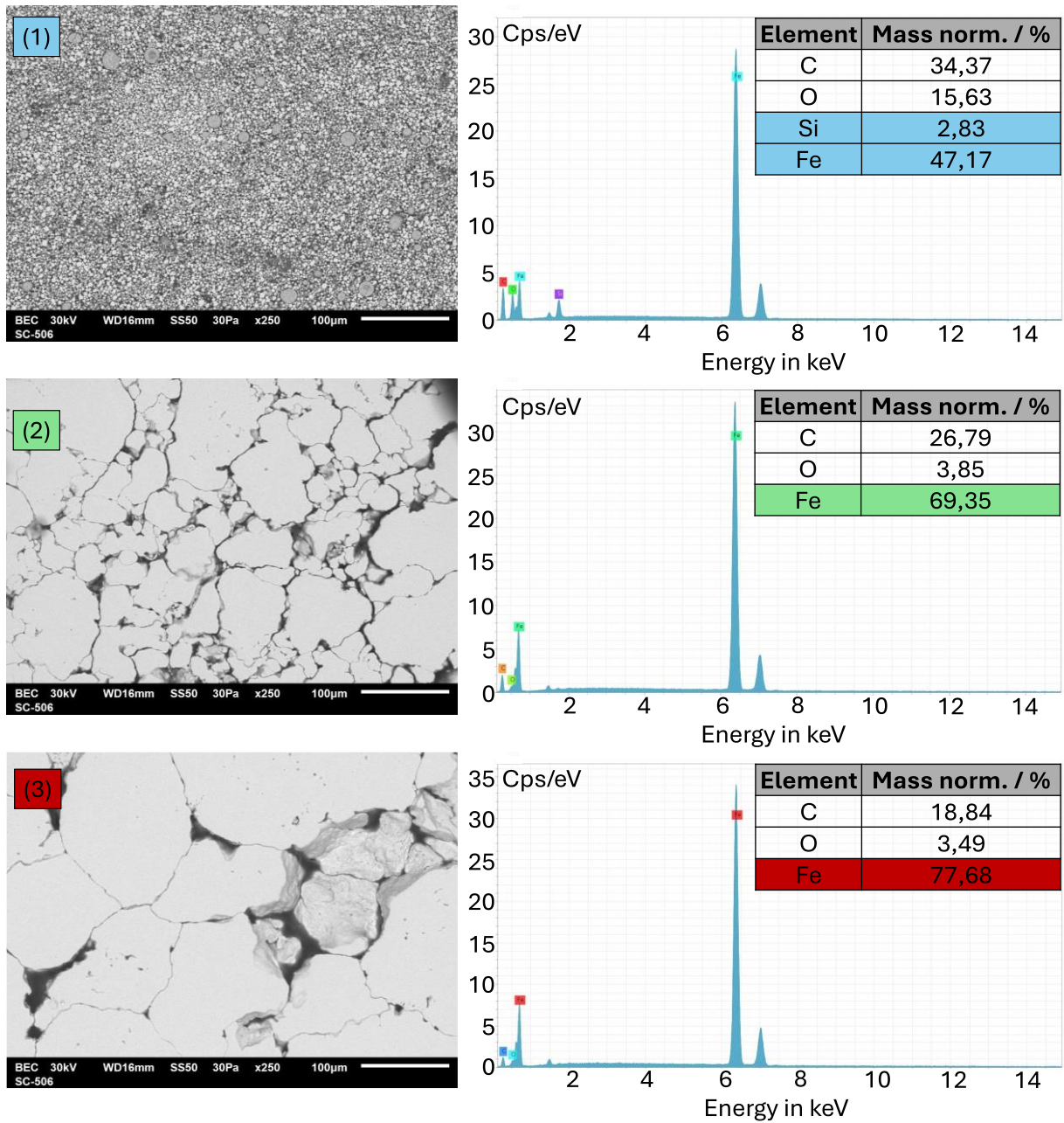

**Figure 4.** Scanning electron microscope (SEM) images (left) and energy-dispersive X-ray (EDX) spectroscopy results (right) of (1) Neosid PFS3 (magnetic particles in polymeric matrix of poly-oxymethylene, which should be responsible for at least some of the carbon and also the oxygen content), (2) Höganäs Somaloy 130i 1P 600 (chemically passivated magnetic particles compacted below sintering), and (3) Höganäs Somaloy Prototype (likewise chemically passivated magnetic particles compacted below sintering).

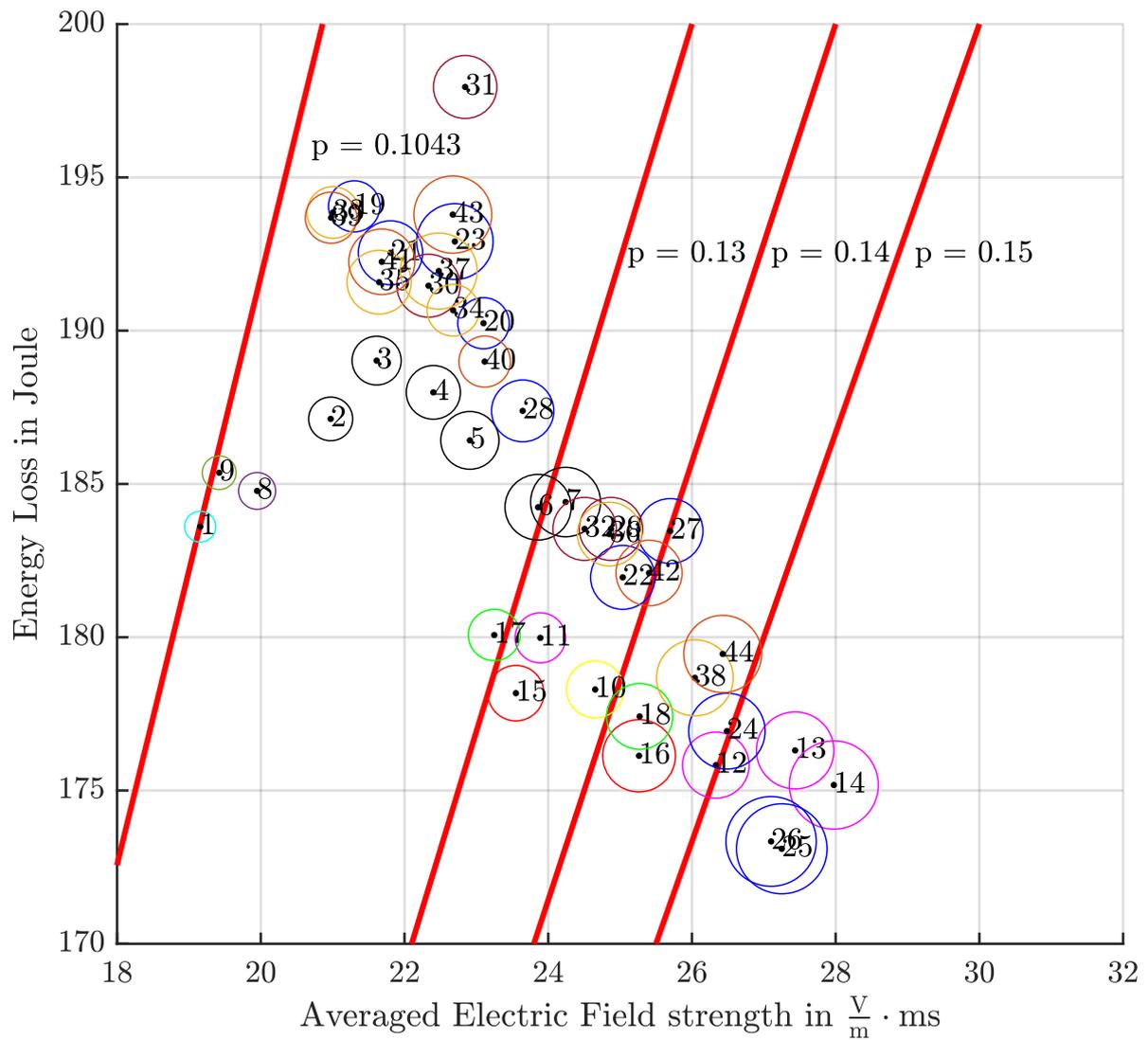

**Figure 5.** Scatter plot of all material configurations and their respective characteristics: averaged electric field strength in the target on the abscissa and energy loss on the ordinate. The color of the circles indicates the specific material; diameter indicates the mass. The red lines constitute constant ratios between energy loss and field strength. The leftmost line resides represents air coils ($p = 0.1043 \, \frac{\text{V} \cdot \text{s}}{\text{J} \cdot \text{m}}$).

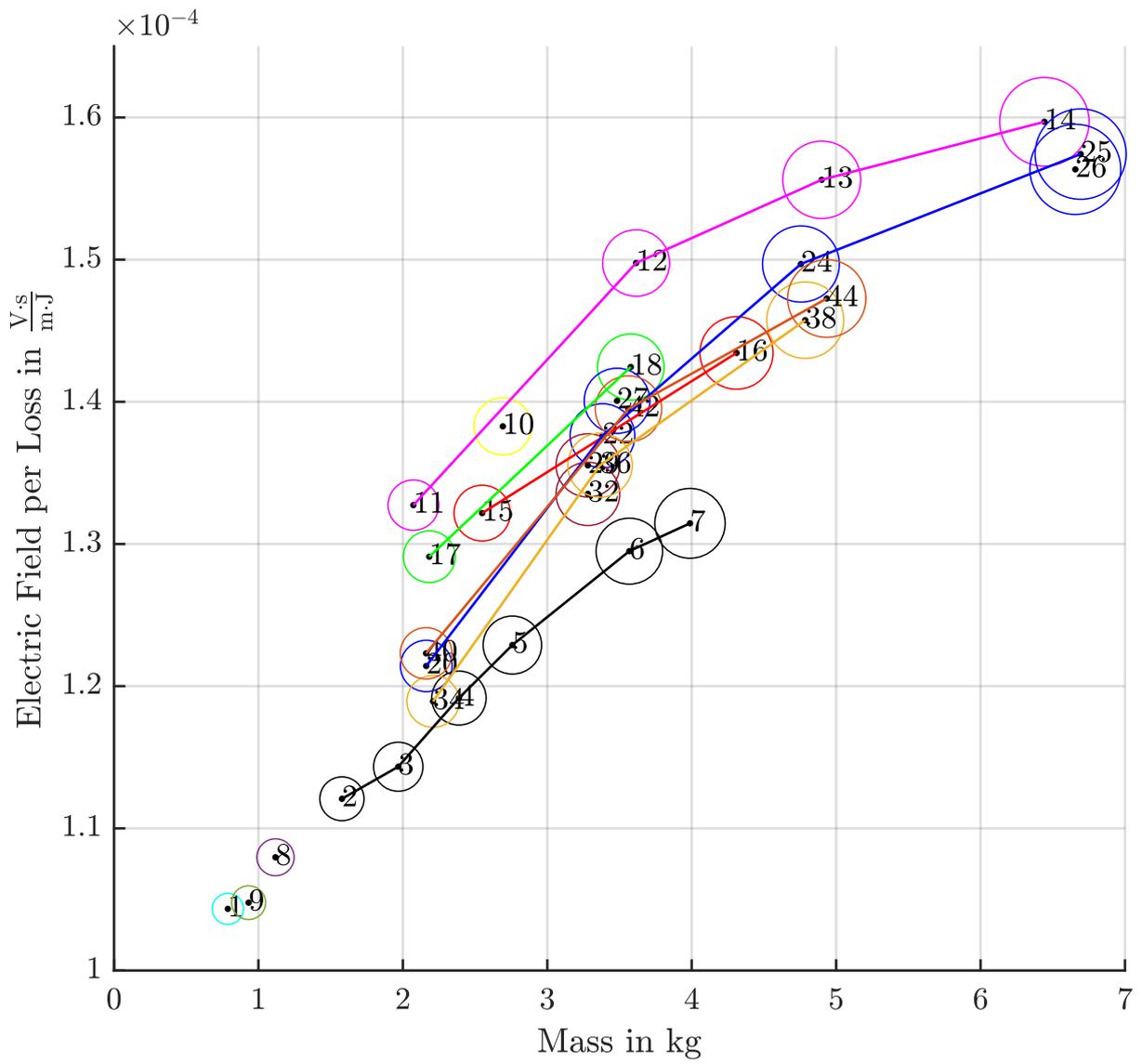

**Figure 6.** Electric field per loss performance index over the total mass of coil and material. Number, color, and size of the circles match Figure 3. The air coil (1) is by far the lightest but also the least performant. Of all the materials, Somaloy Prototype Material scores best in this metric.

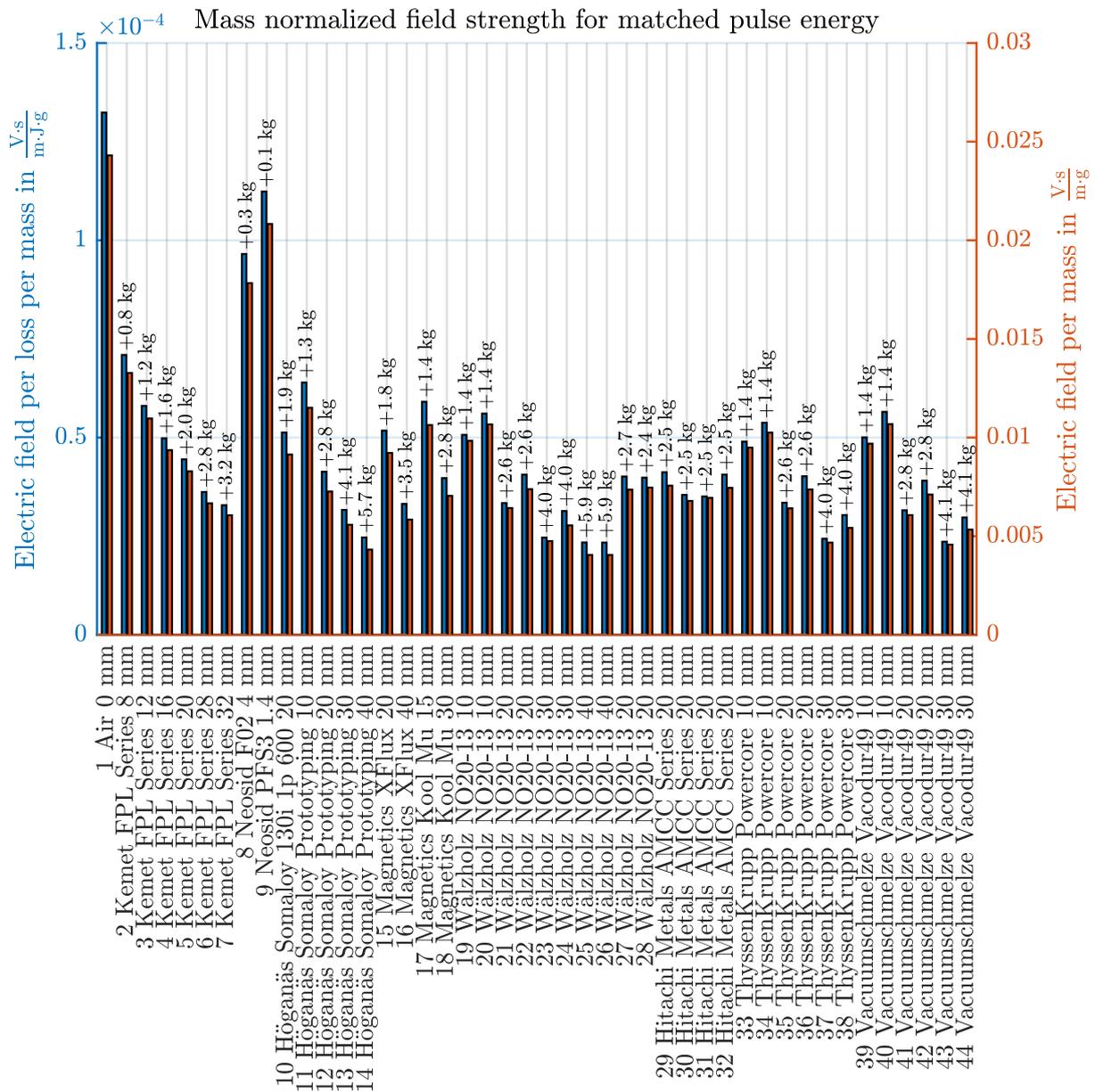

**Figure 7a.** Efficiency (field per loss) and overall target electric field strength normalized to the total mass of the coil including the magnetic material. The bare air coil scores best due to its light weight, whereas no material's field gain or loss reduction (due to smaller currents) can compensate the additional mass (higher is better).

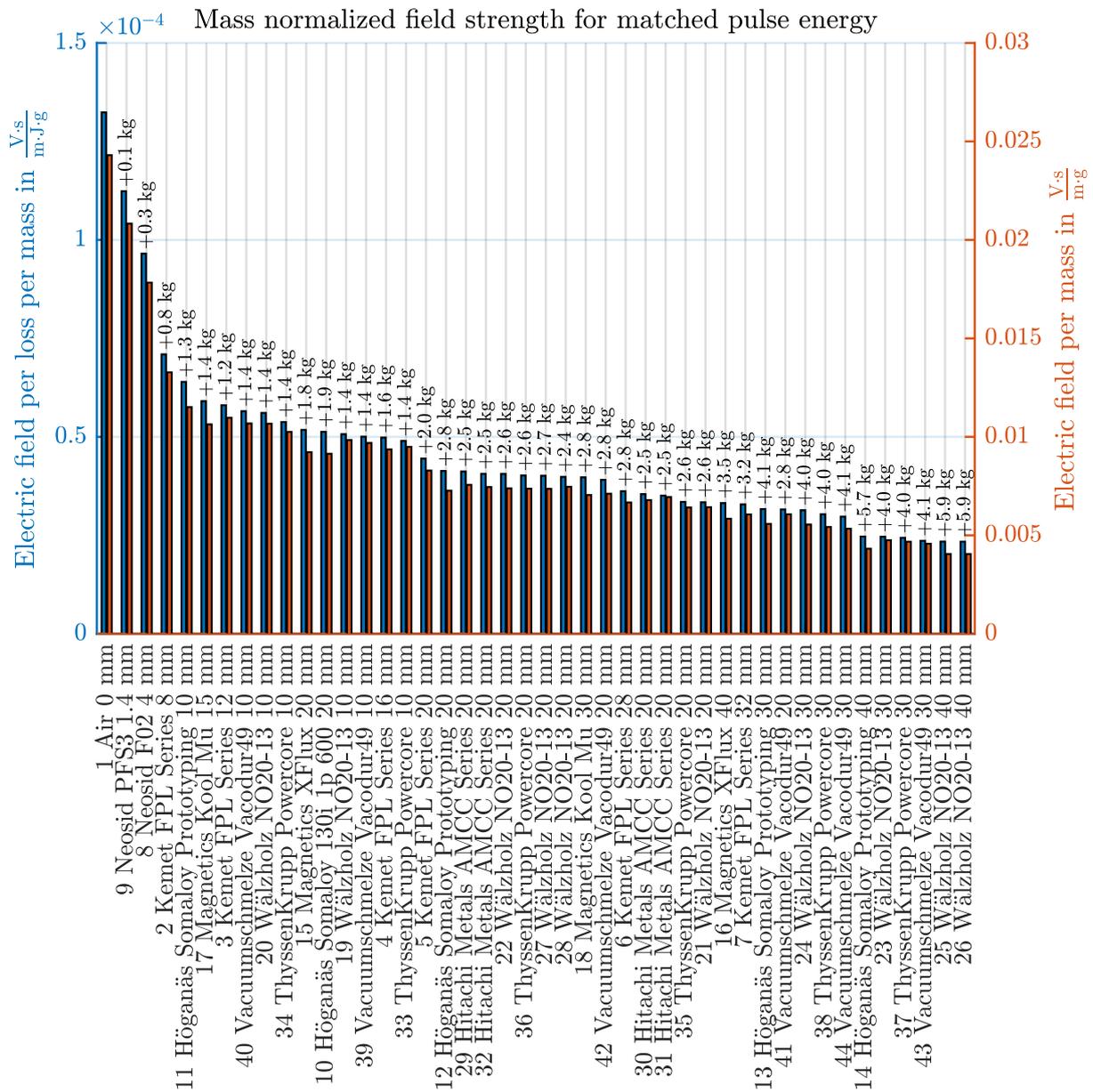

**Figure 7b.** Efficiency (field per loss) and overall target electric field strength normalized to the total mass of the coil including the magnetic material sorted in descending order. The bare air coil scores best due to its light weight, whereas no material's field gain or loss reduction (due to smaller currents) can compensate the additional mass (higher is better).

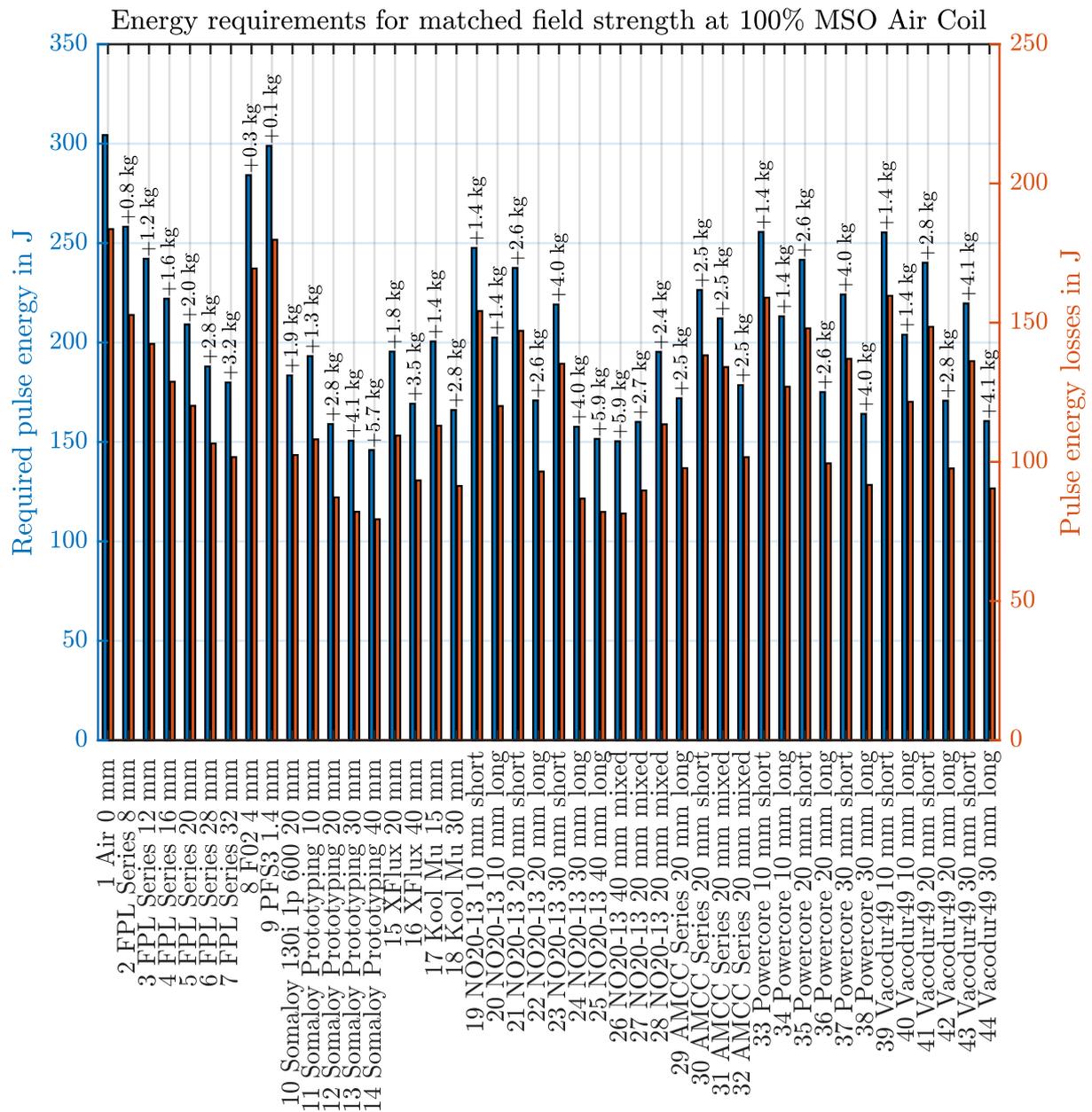

**Figure 8.** Pulse energy and relative losses for every material combination to achieve the electric target field of the reference air coil design at 100 % MSO. The annotation on top of the bars indicates the mass increase on top of the coils base weight (air coil).

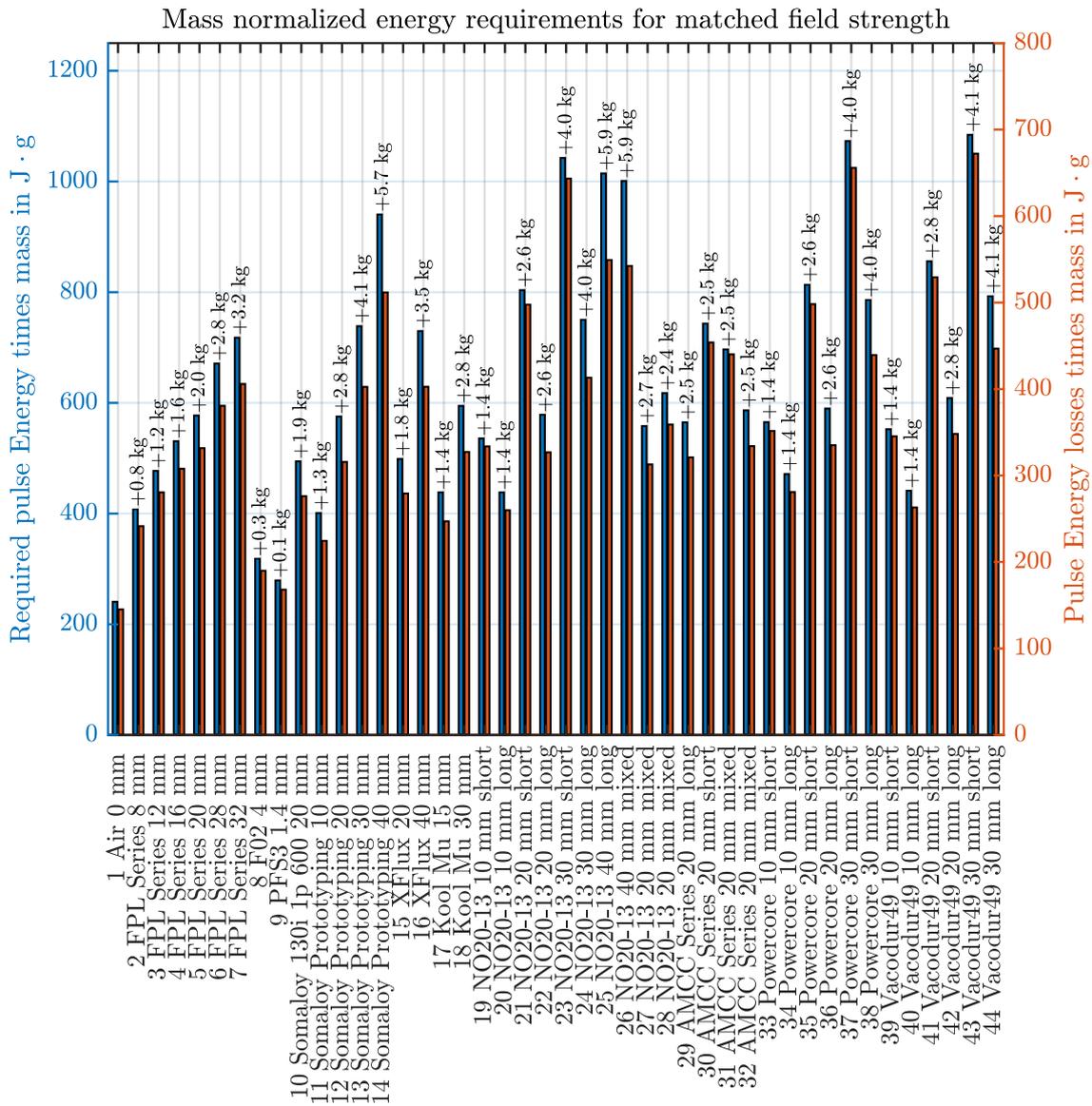

**Figure 9.** Pulse energy and relative losses for every material combination to achieve the electric target field of the reference air coil design at 100 % MSO normalized to the coil weight (lower is better).

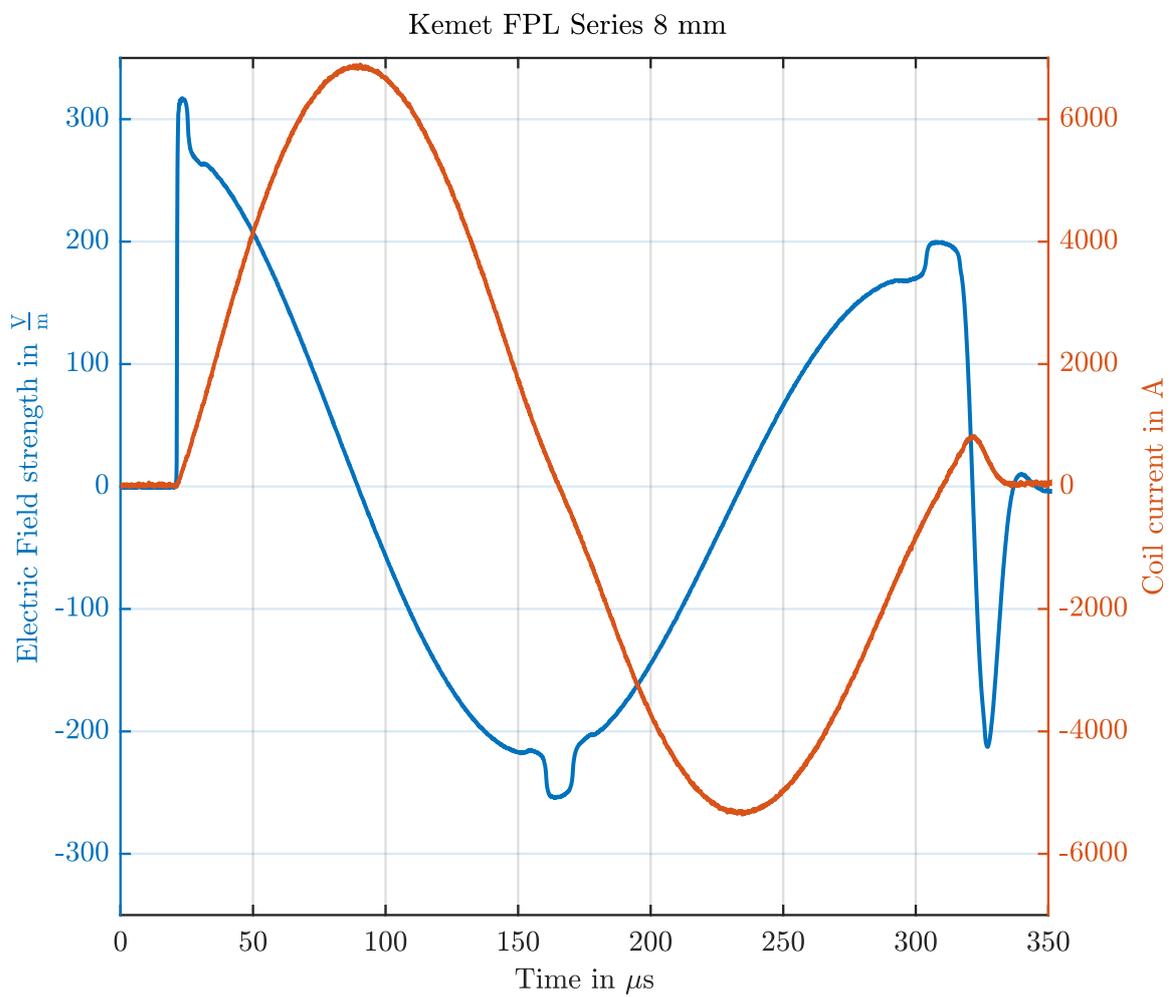

**Figure 10.** Sampled electric field at the target(blue) and coil current (orange) for material sample no. 2 (Kemet FPL 8 mm) at 100% MSO. The artefacts at the maxima of the field, around the minima of the current represent the ranges in which the material is not saturated and amplifies the flux.

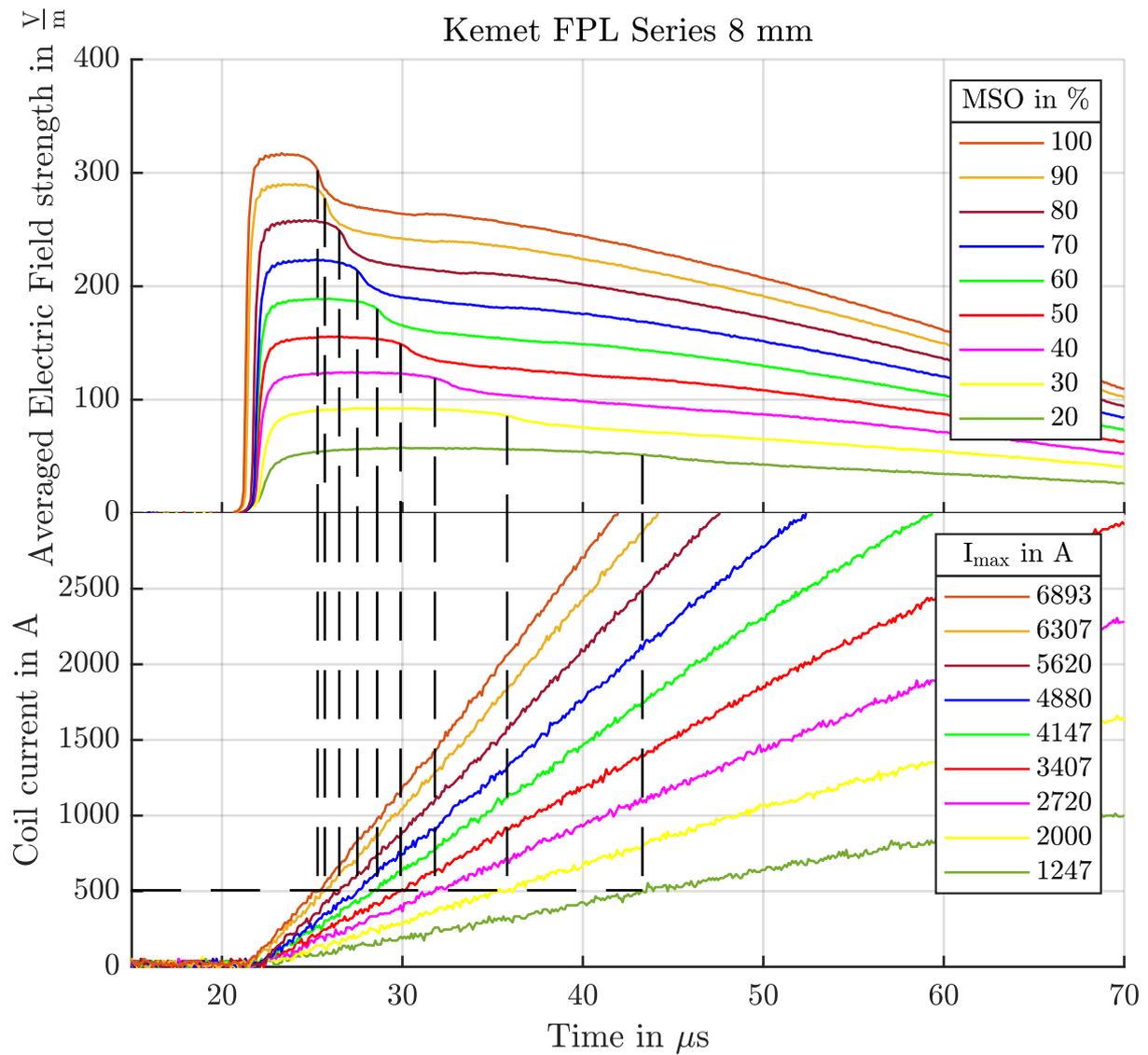

**Figure 11.** Top panel represents the induced electric field strength at the target for the Kemet FPL Series 8 mm material sample (sample no. 2). The lower panel aligns it with the corresponding coil current traces. The induced electric field clearly saturates the applied material very systematically at a current of about 500 A, which represents a certain flux level, in this case far below the sine's peaks (1170 A to 6700 A). At this level, the material's performance approaches air so that flux lines leave the material. Thicker material layers can move this level towards higher currents.

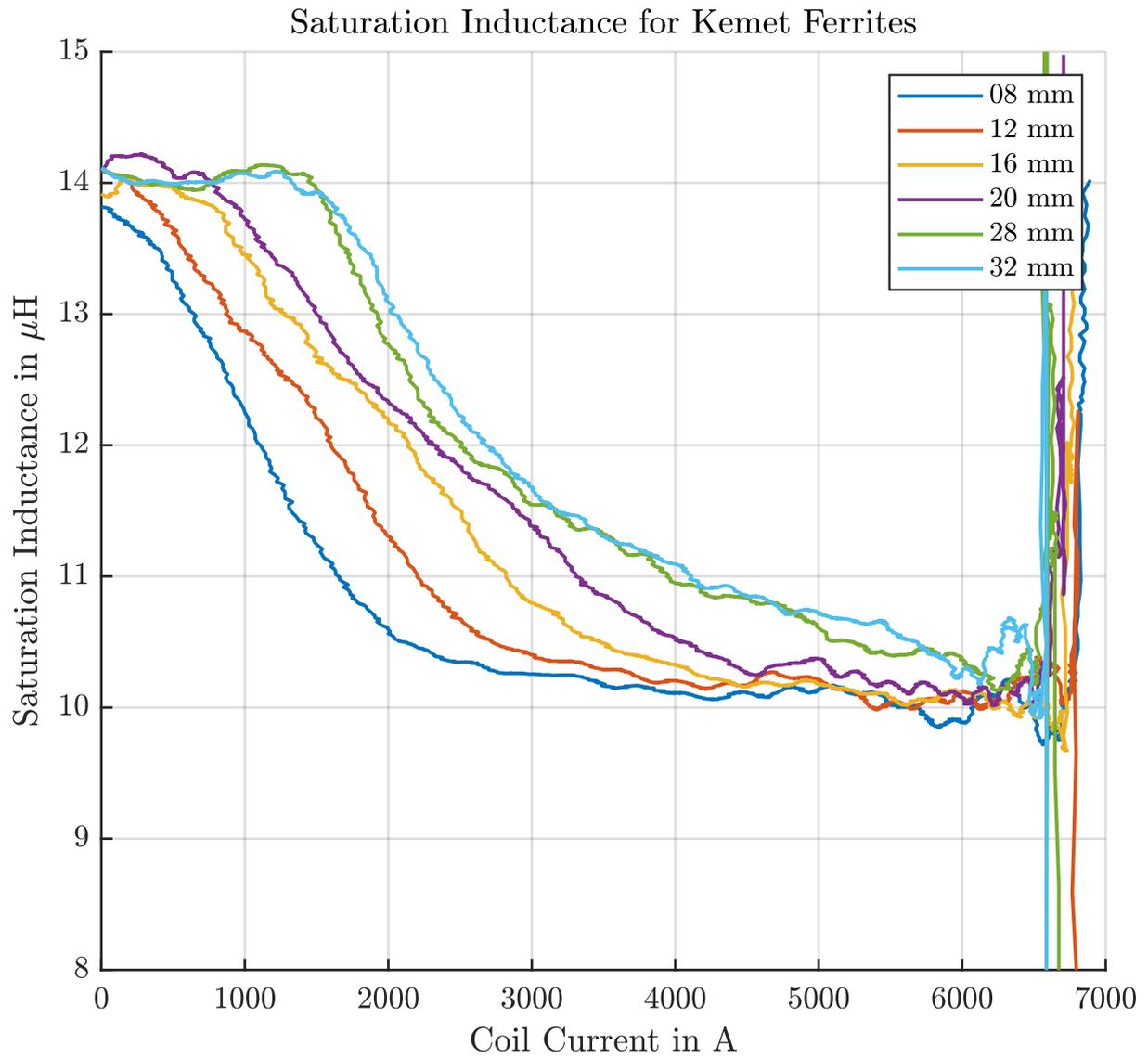

**Figure 12.** Effective inductance over the coil current, which demonstrates how the inductance drops towards the level of the air coil with increasing saturation in the material. In the saturated state, the material maintains little field-amplifying effects but merely adds extra mass and losses in comparison with an air coil.

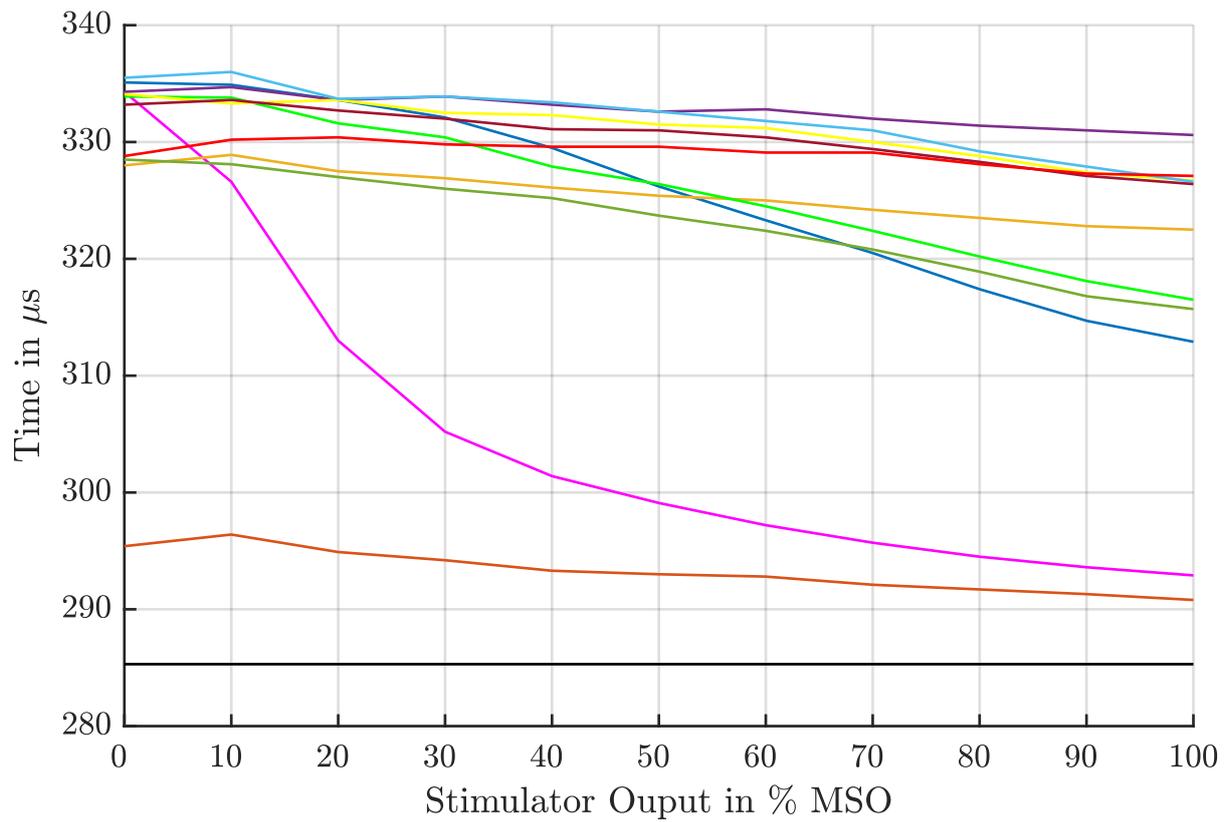

**Figure 13.** Pulse length of the LC oscillator circuit formed by the pulse capacitor and the stimulation coil for various materials and stimulator outputs. The inductance drop in saturation shortens the pulse.

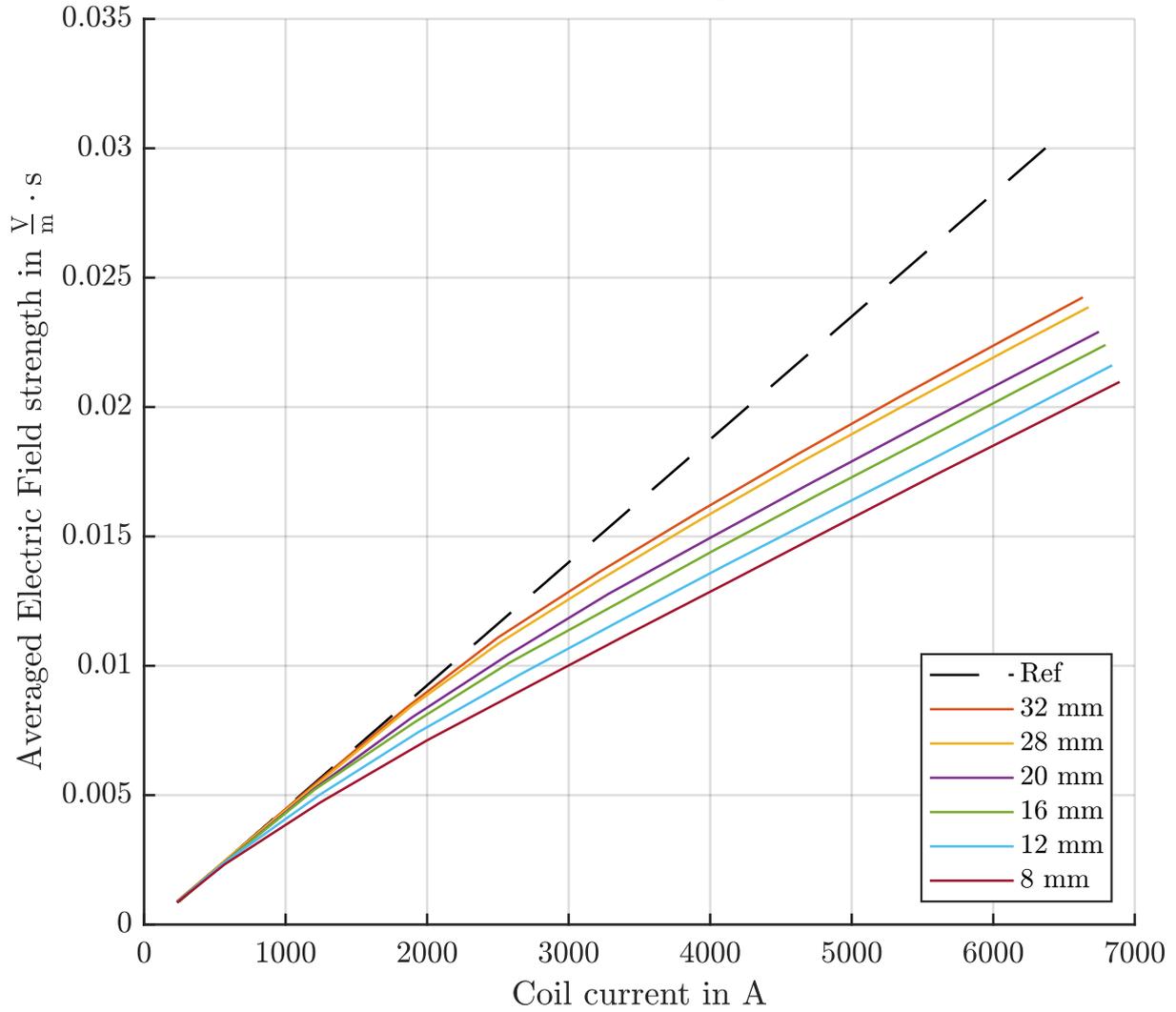

**Figure 14.** Saturation curves for Kemet FPL ferrite materials. The black line extends the initial gradient of the non-saturated materials and thus represents an ideal material without saturation. The colored lines represent the nonlinear materials. As above, thicker material can move the saturation towards higher coil currents.

**Table 1.** Material configurations

| Index | Category | Vendor | Material | Thickness | Weight | Orientation* |
|---|---|---|---|---|---|---|
| 1 | Air | | Air | | 0 g | |
| 2 | Ferrite | Kemet | FPL Series | 8 mm | 790 g | |
| 3 | Ferrite | Kemet | FPL Series | 12 mm | 1180 g | |
| 4 | Ferrite | Kemet | FPL Series | 16 mm | 1600 g | |
| 5 | Ferrite | Kemet | FPL Series | 20 mm | 1970 g | |
| 6 | Ferrite | Kemet | FPL Series | 28 mm | 2780 g | |
| 7 | Ferrite | Kemet | FPL Series | 32 mm | 3200 g | |
| 8 | Ferrite | Neosid | F02 | 4 mm | 330 g | |
| 9 | Ferrite | Neosid | PFS3 | 1.4 mm | 144 g | |
| 10 | Powdercore | Höganäs | Somaloy 130i 1P 600 | 20 mm | 1904 g | |
| 11 | Powdercore | Höganäs | Somaloy Prototyping | 10 mm | 1285 g | |
| 12 | Powdercore | Höganäs | Somaloy Prototyping | 20 mm | 2827 g | |
| 13 | Powdercore | Höganäs | Somaloy Prototyping | 30 mm | 4112 g | |
| 14 | Powdercore | Höganäs | Somaloy Prototyping | 40 mm | 5654 g | |
| 15 | Powdercore | Magnetics | XFlux | 20 mm | 1761 g | |
| 16 | Powdercore | Magnetics | XFlux | 40 mm | 3522 g | |
| 17 | Powdercore | Magnetics | KoolMu | 15 mm | 1395 g | |
| 18 | Powdercore | Magnetics | KoolMu | 30 mm | 2790 g | |
| 19 | Sheet Material | Wälzholz | NO20-13 | 10 mm | 1374 g | short |
| 20 | Sheet Material | Wälzholz | NO20-13 | 10 mm | 1374 g | long |
| 21 | Sheet Material | Wälzholz | NO20-13 | 20 mm | 2594 g | short |
| 22 | Sheet Material | Wälzholz | NO20-13 | 20 mm | 2594 g | long |
| 23 | Sheet Material | Wälzholz | NO20-13 | 30 mm | 3968 g | short |
| 24 | Sheet Material | Wälzholz | NO20-13 | 30 mm | 3968 g | long |
| 25 | Sheet Material | Wälzholz | NO20-13 | 40 mm | 5905 g | long |
| 26 | Sheet Material | Wälzholz | NO20-13 | 40 mm | 5868 g | mixed |
| 27 | Sheet Material | Wälzholz | NO20-13 | 20 mm | 2695 g | mixed |
| 28 | Sheet Material | Wälzholz | NO20-13 | 20 mm | 2372 g | mixed |
| 29 | Sheet Material | Hitachi | AMCC Series | 20 mm | 2493 g | long |
| 30 | Sheet Material | Hitachi | AMCC Series | 20 mm | 2493 g | short |
| 31 | Sheet Material | Hitachi | AMCC Series | 20 mm | 2493 g | mixed |
| 32 | Sheet Material | Hitachi | AMCC Series | 20 mm | 2493 g | mixed |
| 33 | Sheet Material | ThyssenKrupp | Powercore | 10 mm | 1421 g | short |
| 34 | Sheet Material | ThyssenKrupp | Powercore | 10 mm | 1421 g | long |
| 35 | Sheet Material | ThyssenKrupp | Powercore | 20 mm | 2577 g | short |
| 36 | Sheet Material | ThyssenKrupp | Powercore | 20 mm | 2577 g | long |
| 37 | Sheet Material | ThyssenKrupp | Powercore | 30 mm | 3998 g | short |
| 38 | Sheet Material | ThyssenKrupp | Powercore | 30 mm | 3998 g | long |
| 39 | Sheet Material | Vacuumschmelze | Vacodur49 | 10 mm | 1373 g | short |
| 40 | Sheet Material | Vacuumschmelze | Vacodur49 | 10 mm | 1373 g | long |
| 41 | Sheet Material | Vacuumschmelze | Vacodur49 | 20 mm | 2774 g | short |
| 42 | Sheet Material | Vacuumschmelze | Vacodur49 | 20 mm | 2774 g | long |
| 43 | Sheet Material | Vacuumschmelze | Vacodur49 | 30 mm | 4147 g | short |
| 44 | Sheet Material | Vacuumschmelze | Vacodur49 | 30 mm | 4147 g | long |

* The orientation *long* means that each single sheet is aligned perpendicular to the long side of the coil, while *short* is aligned analogously to the short side. The orientation *mixed* describes an individual combination of different lamination directions.

**Table 2.** Categorical overview of the individual materials. Material thickness and lamination directions not included.

| Index | Category | Vendor | Material | Grain size (gs) or sheet thickness (st) | Material constellation | Saturation Flux Density | Mass Density | Relative Permeability |
|---|---|---|---|---|---|---|---|---|
| 1 | Ferrite | Kemet | FPL | n/a | MnZn | 0.52 T | 4.9 g/cm$^3$ | 3000 |
| 2 | Ferrite | Neosid | F02 | n/a | MnZn | 0.34 T | 4.5 g/cm$^3$ | 1800 |
| 3 | Powder core | Neosid | PFS3 | gs: ≤ 10 µm | FeSi POM Binder | 1 T | 4.8 g/cm$^3$ | 12 |
| 4 | Powder core | Höganäs | Somaloy 130i 1P 600 | gs: 20 … 100 µm | Fe | 2.4 T | 7.22 g/cm$^3$ | 290 |
| 5 | Powder core | Höganäs | Somaloy Prototype | gs: ~100 µm | Fe | 2.3 T | 7.3 g/cm$^3$ | 430 |
| 6 | Powder core | Magnetics | XFlux | unspecified | FeSi | 1.6 T | 7.6 g/cm$^3$ | 26 |
| 7 | Powder core | Magnetics | Kool Mu | unspecified | FeSiAl | 1 T | 5.8 g/cm$^3$ | 90 |
| 8 | Sheet material | Wälzholz | NO20-13 | st: 200 µm | 94-97 % Fe 3-6 % Si | 1.59 T | 7.68 g/cm$^3$ | 20000 |
| 9 | Sheet material | Hitachi Metals | AMCC Series | st: 25 µm | 85-95 % Fe 5-10 % Si 1-5 % B | 1.56 T | 7.18 g/cm$^3$ | 45000 |
| 10 | Sheet material | ThyssenKrupp | Powercore (KO) | st: 230 µm | 96–97 % Fe 3-4 % Si | 2.03 T | 7.65 g/cm$^3$ | 20000 |
| 11 | Sheet material | Vacuumschmelze | Vacodur49 | st: 357 µm | 49 % Fe 49 % Co 2 % Va | 2.3 T | 8.12 g/cm$^3$ | 15000 |